\documentclass[aps,prl,twocolumn,superscriptaddress]{revtex4}

\usepackage{xcolor}
\usepackage{amsmath,bm}
\usepackage{graphicx}
\usepackage{upgreek}

\newcommand{\amin}{\alpha_\mathrm{min}}

\newcommand{\ADEV}{\Delta Y_{\mathrm{Allan}}}
\newcommand{\mean}[1]{\left\langle #1 \right\rangle}

\newcommand{\vecro}{\mathbf{r_0}}
\newcommand{\vecBrf}{\mathbf{B_{\mathrm{rf}}}}
\newcommand{\vecBec}{\mathbf{B_{\mathrm{ec}}}}

\newcommand{\Usat}{U_{\mathrm{sat}}}
\newcommand{\Bsat}{B_{\mathrm{sat}}}
\newcommand{\Bec}{B_{\mathrm{ec}}}

\newcommand{\hatx}{\mathbf{\widehat{x}}}
\newcommand{\haty}{\mathbf{\widehat{y}}}

\newcommand{\Jmax}{J_{\mathrm{max}}}
\newcommand{\vecJmax}{\mathbf{J_{\mathrm{max}}}}
\newcommand{\Gdark}{\Gamma_{\mathrm{dark}}}
\newcommand{\Gpump}{\Gamma_{\mathrm{p}}}
\newcommand{\Gprobe}{\Gamma_{\mathrm{pr}}}

\newcommand{\lr}[1]{\left(  #1 \right)}

\newcommand{\abs}[1]{\left|  #1 \right|}
\newcommand{\vecB}{\mathbf{B}}
\newcommand{\vecJ}{\mathbf{J}}
\newcommand{\LR}[1]{\left[  #1 \right]}
\begin{document}

\title{Detection of low-conductivity objects using eddy current measurements with an optical magnetometer} 

\author{Kasper Jensen}
\thanks{corresponding author\\ 
email: Kasper.Jensen@nottingham.ac.uk}
\affiliation{School of Physics and Astronomy, University of Nottingham, University Park, Nottingham NG7 2RD, England, United Kingdom}
\affiliation{Niels Bohr Institute, University of Copenhagen, Blegdamsvej 17, 2100 Copenhagen, Denmark}

\author{Michael Zugenmaier}
\affiliation{Niels Bohr Institute, University of Copenhagen, Blegdamsvej 17, 2100 Copenhagen, Denmark}

\author{Jens Arnbak}
\affiliation{Niels Bohr Institute, University of Copenhagen, Blegdamsvej 17, 2100 Copenhagen, Denmark}

\author{Hans St{\ae}rkind}
\affiliation{Niels Bohr Institute, University of Copenhagen, Blegdamsvej 17, 2100 Copenhagen, Denmark}

\author{Mikhail~V.~Balabas}
\affiliation{Niels Bohr Institute, University of Copenhagen, Blegdamsvej 17, 2100 Copenhagen, Denmark}
\affiliation{Department of Physics, St Petersburg State University, Universitetskii pr. 28, 198504 Staryi Peterhof, Russia}

\author{Eugene S. Polzik}
\affiliation{Niels Bohr Institute, University of Copenhagen, Blegdamsvej 17, 2100 Copenhagen, Denmark}

%%%%%%%%%%%%%%%%%%%%%%%%%%%%%%%%%%%%%%%%%%%%%%%%%%%%%%%

\begin{abstract}
Detection and imaging of an electrically conductive object at a distance can be achieved by inducing eddy currents in it and measuring the associated magnetic field. 
We have detected low-conductivity objects  
with an optical magnetometer based on room-temperature cesium atomic vapor and a noise-canceling differential technique which increased the signal-to-noise ratio (SNR) by more than three orders of magnitude. 
We detected small containers with a few mL of salt-water with conductivity ranging from 4--24~S/m with a good SNR.
This demonstrates that our optical magnetometer should be capable of detecting objects with conductivity~$<1$~S/m with a SNR $>1$
and opens up new avenues for using optical magnetometers to image low-conductivity biological tissue including the human heart which would enable non-invasive diagnostics of heart diseases.
\end{abstract}

\maketitle

Optical magnetometers \cite{Budker07} based on laser-interrogation of cesium or rubidium vapor  can detect magnetic fields with sub-fT/$\sqrt{\rm{Hz}}$ sensitivity \cite{Kominis2003nature,Savukov05prl,Wasilewski2010prl,Chalupczak2012apl}. This high sensitivity is particularly useful for biomedical applications where tiny magnetic fields from the human body are detected.
For example, optical magnetometers have detected brain activity \cite{Xia2006,Boto2017neuroimage,Boto2018nature}, the heartbeat from adults \cite{Bison2009apl} and fetuses \cite{Wyllie2012ol,Alem2015PhysMedBio}, and nerve impulses \cite{Jensen2016scirep}.
Optical magnetometers can potentially also be used to non-invasively image the electrical conductivity $\sigma$ of the heart \cite{Marmugi2016scirep} using a technique called magnetic induction tomography (MIT) \cite{Griffiths1999,Griffiths2001}. 
In MIT of the heart, one or more coils are used to induce eddy currents in the heart and an image of the heart is constructed from measurements of the associated induced magnetic field. This is a challenging task for several reasons, with the main one being the low conductivity $\sigma \lesssim 1$~S/m of the heart \cite{Marmugi2016scirep}.

Imaging of low-conductivity objects has previously been done using coils for inducing and detecting the eddy currents.
Large containers ($\approx 500$~mL) with salt-water with conductivity as low as 0.7~S/m has been imaged  \cite{Griffiths1999,Korjenevsky2000,Wei2012}, and more recently, the spinal column has been imaged with a single scanning coil \cite{Feldkamp2015}. 
Optical magnetometers have several advantages compared to induction coils, in particular, they are widely tunable and  can achieve high sensitivity which is fundamentally independent of the operating frequency. This is in contrast to  induction coils which are sensitive to the change in magnetic flux and therefore have worse sensitivity the lower the frequency. So far, optical magnetometers have been used to image highly conductive metallic samples 
($\sigma \approx 10^6 \text{--} 10^8$~S/m) \cite{Wickenbrock2014ol,Wickenbrock2016apl,Deans2016apl} and also recently semiconductor materials ($\sigma = 500\text{--}10.000$~S/m) \cite{Marmugi2018arxiv}.

In this work, we introduce a differential technique which improves the signal-to-noise ratio by more than three orders of magnitude and then
demonstrate detection of small containers with 8~mL of salt-water with conductivity as low as 4~S/m. 
This represents an improvement by two orders of magnitude compared to previous results with optical magnetometers \cite{Marmugi2018arxiv} and is a big step towards magnetic induction tomography of biological tissue with optical magnetometers. 

\begin{figure}[th]
\centering
\includegraphics[width=0.48\textwidth]{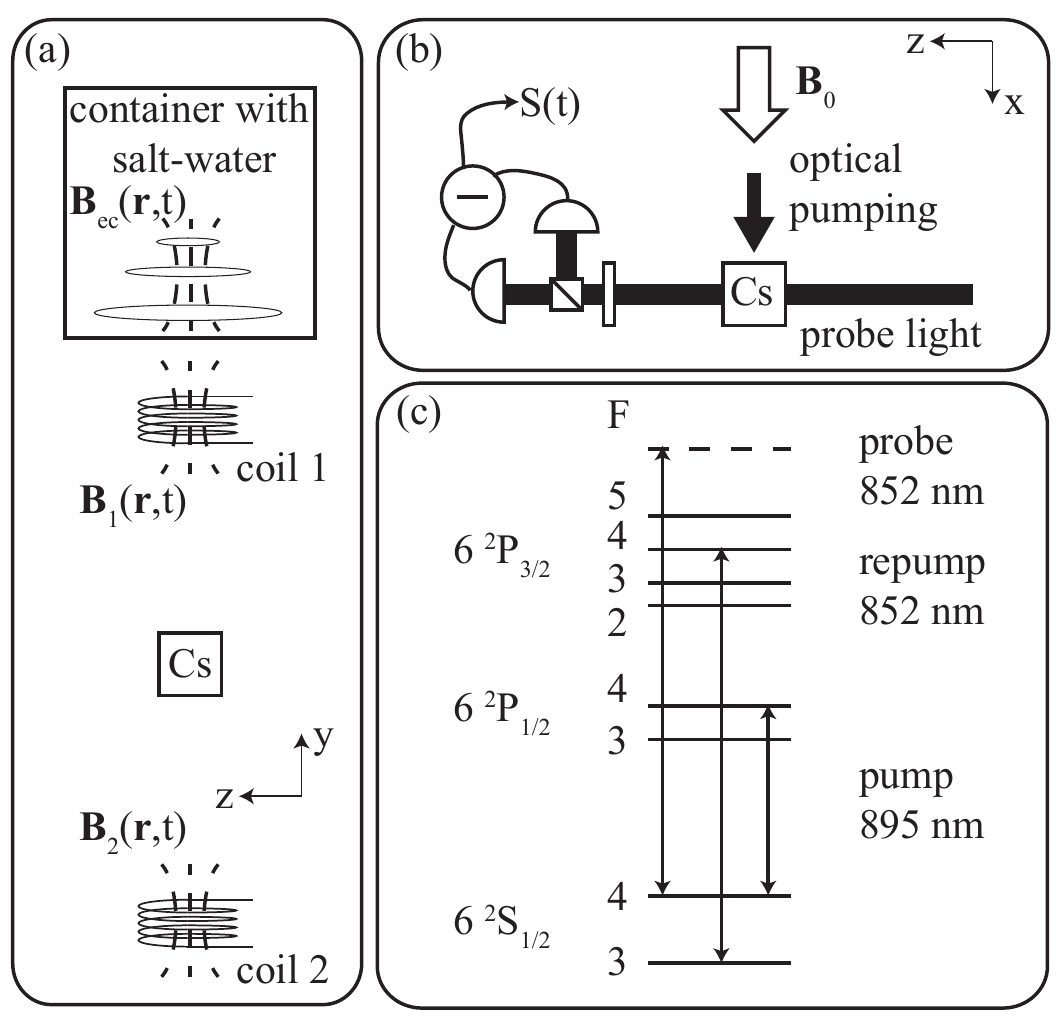}
\caption{
(a) Setup for detecting eddy currents. 
(b) Optical pumping and probing of the cesium atomic spins. 
(c) Cesium level scheme and laser wavelengths. 
The probe light is 1.6~GHz detuned.}
\label{fig:setup}
\end{figure}

We first discuss the standard approach for detecting and imaging a conductive object, in our case a container with salt-water. Later we will discuss the differential technique. Consider a conductive object, a magnetometer, and a coil [denoted coil 1 in Fig.~\ref{fig:setup}(a)] that generates a primary magnetic field 
$\mathbf{B_1}(\mathbf{r}, t)$ oscillating at the frequency $\omega=2\pi\nu$. The primary field induces eddy currents in the object which in turn generate a secondary magnetic field $\vecBec(\mathbf{r}, t)$. 
One can measure the total field $\mathbf{B_1}(\mathbf{r},t) + \vecBec(\mathbf{r},t)$ and by scanning the magnetometer or the object around it is possible to construct an image of the conductivity \cite{Wickenbrock2014ol,Wickenbrock2016apl,Deans2016apl,Marmugi2018arxiv}. Varying the frequency $\omega$ can be useful for 3D imaging \cite{Marmugi2016scirep} and for material characterization \cite{Wickenbrock2016apl,Marmugi2018arxiv}. 

It is instructive to note that the primary field is attenuated while penetrating into the object due to the skin effect. The skin depth is
$ \delta(\omega) \approx \sqrt{2/\lr{\omega \mu_0 \sigma}}$,  where $\mu_0$ is the vacuum permeability and we assumed that the object is non-magnetic.
When the thickness $t$ of the object is much smaller than the skin depth $t \ll \delta(\omega)$, 
the secondary field is $90^\circ$ out of phase with the primary field and the ratio $\alpha$ of the amplitude $\Bec(\vecro)$ of the secondary field to the amplitude of the primary field $B_1(\vecro)$ at the magnetometer position $\vecro$ is \cite{Griffiths1999, Griffiths2007,Sup}
\begin{equation}
\alpha \equiv \Bec(\vecro)/B_1(\vecro) 
\approx  -A \sigma \omega \mu_0
\approx  -2 A /\LR{\delta(\omega)}^2,
\label{eq:BecB1}
\end{equation}
where $A$ is a geometrical factor with dimensions of length squared. 
For a  $\lr{2~\mathrm{cm}}^3$ container  with salt-water with conductivity $\sigma=10.7$~S/m we calculate  $\delta = 11$~cm and estimate 
$\abs{\alpha} \approx 1.5 \cdot 10^{-4}$ \cite{Sup} when the frequency is
2~MHz. 
We demonstrate that it is possible to detect such a small change in signal with an optical magnetometer when using a differential technique.

The key component of our magnetometer is a paraffin-coated  cesium vapor cell with a $\lr{5~\mathrm{mm}}^3$ inner volume \cite{Jensen2018scirep}. The cesium atoms are spin-polarized in the $x$-direction using circularly polarized pump and repump light  and are detected using linearly polarized probe light [see Fig.~\ref{fig:setup}(b)--(c)]. 
We denote the total angular momentum in the $F=4$ hyperfine ground state manifold $\vecJ$ and full polarization corresponds to $ \vecJ = \vecJmax  = 4 N_A \hatx$, where $N_A$ is the number of cesium atoms. The atoms are placed in a static magnetic field 
$B_0 \hatx$ and we are interested in detecting an oscillating magnetic field 
$\vecBrf(t) = \LR{B_c \cos \lr{\omega t} + B_s \sin \lr{\omega t}} \haty$. 
The time evolution of the atomic spins is modelled using the differential equation \cite{Ledbetter2008pra}
\begin{equation}
\frac{d \vecJ}{dt} = \gamma \vecJ \times \vecB +
\Gpump \vecJmax -
\lr{\Gpump + \Gprobe + \Gdark }\vecJ ,
\end{equation}
where $\gamma$ is the cesium gyromagnetic ratio,
$\vecB =B_0 \hatx + \vecBrf(t)$,
$\Gpump$ is the rate of optical pumping,
$\Gdark$ is the decay rate in the absence of light, and $\Gprobe$ is the decay rate due to the probe light.
We solve the  differential equation in the frame rotating around the $x$-axis at the frequency $\omega$. Denoting the spin-vector  in the rotating frame $\vecJ'$ and assuming a steady state $\frac{d \vecJ'}{dt} = 0$, we find the spin-components 
\begin{eqnarray}
J_x^{'ss} &=& \hphantom{-} J_{ss} 
\frac{\Delta^2+\lr{\delta \omega}^2}{\Delta^2+\lr{\delta \omega}^2 + \gamma^2 \lr{B_c^2+B_s^2}/4 }, \label{eq:Jxprime}\\
J_y^{'ss} &=& -J_{ss} 
\frac{\gamma \lr{B_c \Delta + B_s \delta \omega}/2}{\Delta^2+\lr{\delta \omega}^2 + \gamma^2\lr{B_c^2+B_s^2}/4 } ,\label{eq:Jyprime}\\
J_z^{'ss} &=& \hphantom{-} J_{ss} 
\frac{\gamma \lr{B_s \Delta - B_c \delta \omega}/2}{\Delta^2+\lr{\delta \omega}^2 + \gamma^2 \lr{B_c^2+B_s^2}/4}. \label{eq:Jzprime}
\end{eqnarray}
Here $\delta \omega=\Gpump + \Gprobe + \Gdark$,
$\Delta=\omega-\omega_L$ is the detuning of the applied frequency from the Larmor frequency $\omega_L =\gamma B_0$,
 and $J_{ss} = \Jmax \Gpump/\lr{\Gpump+\Gprobe+\Gdark}$.

\begin{figure*}[th]
\centering
\includegraphics[width=0.9\textwidth]{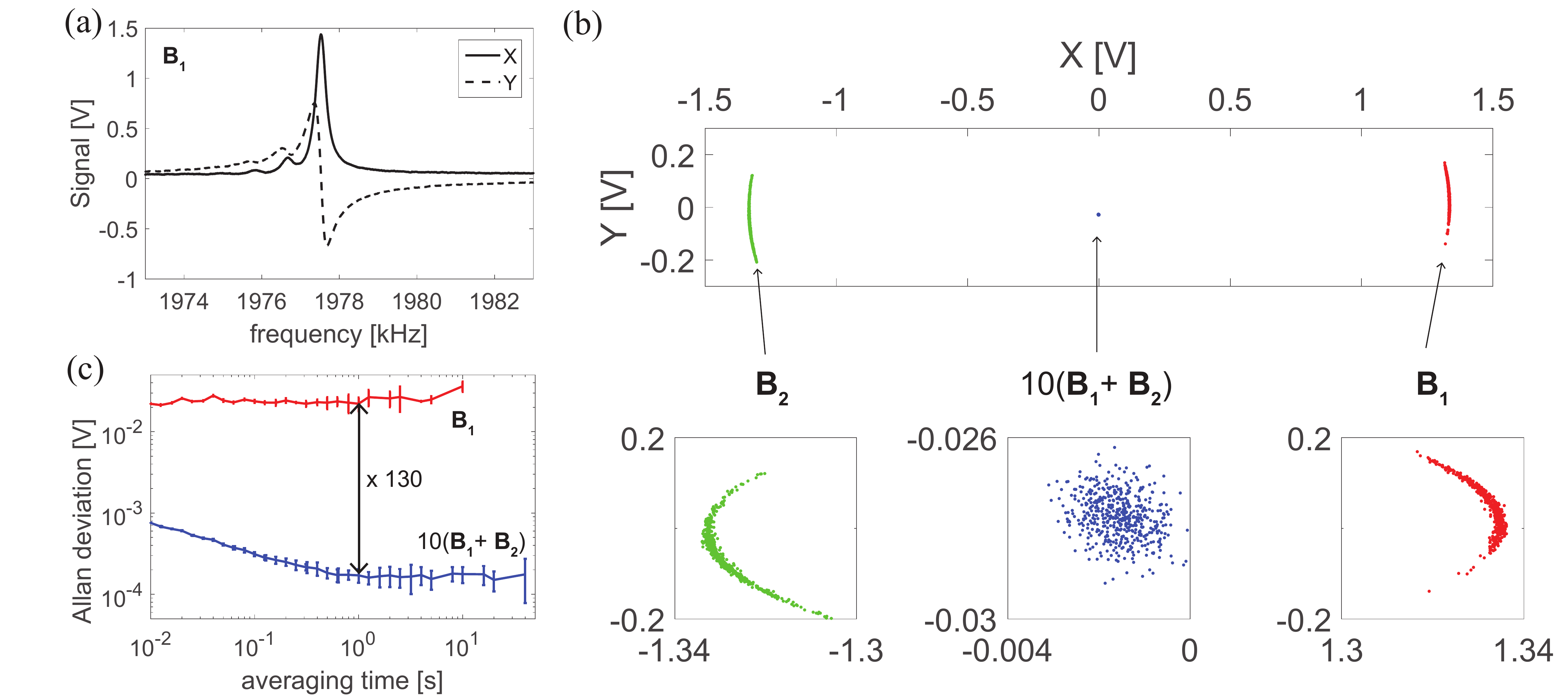}
\caption{(a) Lock-in outputs $X$ and $Y$ as a function of the frequency of the oscillating magnetic field $\mathbf{B_1}$. 
(b) Lock-in outputs when $\omega=\omega_L$. Each data point was integrated for 40~ms. Three data sets each with 500 points are shown: one where $\mathbf{B_1}$ was applied, one where $\mathbf{B_2}$ was applied and one where 
$10\lr{\mathbf{B_1}+\mathbf{B_2}}$ was applied. 
(c) Allan deviation of the $Y$-output when $\mathbf{B_1}$ was applied (top trace) and when $10\lr{\mathbf{B_1}+\mathbf{B_2}}$ was applied (bottom trace). }
\label{fig:calib}
\end{figure*}

If we only consider $B_c$ (i.e. $B_s=0$), we see that 
 $J_y^{'ss}$ and $J_z^{'ss}$ have dispersive and lorentzian lineshapes, respectively, as a function of detuning.
The total width of the resonance is 
$\delta  \omega \cdot  \sqrt{1 + \LR{B_c/\Bsat}^2 }$ 
where $\Bsat \equiv 2 \delta \omega /\gamma$.
This means that the  resonance is power-broadened by the oscillating magnetic field $B_c$.
If the  magnetic field is on resonance ($\Delta  = 0$) we have 
$J_z^{'ss} \propto B_c/\lr{1+\LR{B_c/\Bsat}^2}$ which means that $J_z^{'ss}$ is only linear with the magnetic field for small fields $\abs{B_c} \ll \Bsat$. 

The atoms are probed with linearly polarized light which due to the Faraday effect is rotated by an amount proportional to the spin-component along the probe propagation direction. The light polarization rotation is measured with a balanced detection scheme leading to the magnetometer signal 
\begin{equation}
S(t) \propto J_z(t) = 
\sin \lr{\omega t} J_y'(t) + 
\cos \lr{\omega t} J_z'(t) .
\end{equation}
The rotating spin-components $J_y'$ and $J_z'$ are extracted from the magnetometer signal using lock-in detection at the frequency $\omega$. 
The lock-in provides an in-phase output $X   \propto J_z'$ and an out-of-phase output $Y \propto J_y'$.

We characterize the magnetometer (without any conductive object) by applying the magnetic field 
$\vecBrf(t) = \mathbf{B_1}(\vecro,t) \equiv B_1(\vecro) \cos \lr{\omega t} \haty$.
Figure~\ref{fig:calib}(a) shows the lock-in outputs as a function of frequency. The $X$- and $Y$-outputs have  lorentzian and dispersive lineshapes centered around the Larmor frequency $\nu_L \approx 1978$~kHz as expected from 
Eqs.~(\ref{eq:Jyprime}) and (\ref{eq:Jzprime}) when $B_c=B_1(\vecro)$ and $B_s=0$. 
The small side resonances (towards lower frequencies) are due to the non-linear Zeeman effect \cite{NonlinearZeeman,Julsgaard2004mors}.
The dataset labeled ``$\mathbf{B_1}$" in Fig.~\ref{fig:calib}(b) shows the lock-in outputs when the oscillating magnetic field $\mathbf{B_1}$ is on resonance ($\Delta = 0$).
We see that the mean values are $\mean{X}=1.33$~V and $\mean{Y}\approx 0$  and that there is a significant amount of noise in the $Y$-output.
In order to characterize the noise we calculate the Allan deviation \cite{Allan1987} of the $Y$-output 
which is roughly independent of averaging time with the value $\ADEV = 22$~mV [see Fig.~\ref{fig:calib}(c)].
The noise is mainly due to temporal fluctuations in the $B_0$-field.
A change $\Delta B_0$ in the $B_0$-field shifts the Larmor frequency which  then changes the $Y$-output. Close to the Larmor frequency 
$Y\approx a\cdot\lr{\nu-\nu_L}$ where $a=-7.8$~V/kHz [see Fig.~\ref{fig:calib}(a)] 
which means that a small change of the $Y$-output of 22~mV corresponds to a shift in the Larmor frequency of 2.8~Hz and a relative change in the 
$B_0$-field of $\Delta B_0/B_0 =1.4\cdot 10^{-6}$. 
This small number illustrates that an optical magnetometer requires a very stable $B_0$-field in order to precisely measure an oscillating magnetic field.

When detecting conductive objects, the amplitude of the secondary field is often  much smaller than the amplitude of the primary field. This is the case if the object is much thinner than the skin depth $t \ll \delta(\omega)$
or if the object is far away from the coil or magnetometer. 
For a thin sample, the secondary field is $90^\circ$ out of phase with the primary field such that $\vecBec(\vecro, t) = \Bec(\vecro) \sin \lr{\omega t} \haty$ with $\Bec(\vecro) = \alpha B_1(\vecro)$ where $\abs{\alpha} \ll 1$.
When detecting the total field $\vecBrf(t) = \mathbf{B_1}(\vecro, t) + \vecBec(\vecro, t)$, 
the field from the eddy currents gives a signal in the $Y$-output and the primary field gives a signal in the $X$-output 
[see Eqs.~(\ref{eq:Jyprime}) and (\ref{eq:Jzprime}) with $\Delta =0$, $B_c=B_1(\vecro)$ and $B_s = \Bec(\vecro)$].
It is problematic to detect the total field for several reasons.
First of all, one would like to increase the amplitude of the primary field  as $\Bec(\vecro) \propto B_1(\vecro)$. 
However, when $\abs{B_1(\vecro)} \gtrsim \Bsat$ there is significant power broadening which leads to reduced signal size and non-linearities.
Even if  $\abs{B_1(\vecro)} \ll \Bsat$ such that the magnetometer signal is linear and the lock-in outputs are
 $\mean{X} \propto  B_1(\vecro)$ and 
$\mean{Y} \propto  \Bec(\vecro)=\alpha B_1(\vecro)$,  it is still problematic to measure the total field as in most cases  both the signal and the noise in the magnetometer are proportional to the amplitude of the total signal. 
In particular, if the dominant source of noise is the instability in the $B_0$-field, then both signal and noise are proportional to $B_1(\vecro)$.
If we detect the total field, the  smallest detectable field from the eddy currents is $\Bec(\vecro) = \amin  B_1(\vecro)$ with 
$\abs{\amin} \approx \ADEV/\mean{X}=1.7\cdot 10^{-2}$.
This is clearly not sufficient to detect low conductivity objects such as biological tissue or salt-water phantoms.

In order to mitigate the above mentioned problems, we introduce a differential technique  where we use a second coil [denoted coil 2 in Fig.~\ref{fig:setup}(a)] that generates a magnetic field $\mathbf{B_2}(\vecro, t)$ such that in the absence of the conductive object, the total magnetic field $ \mathbf{B_1}(\vecro, t) +  \mathbf{B_2}(\vecro, t) \approx 0$ at the position of the vapor cell. 
Coil 2 is placed further away from the conductive object than coil 1 such that eddy currents are mainly generated by coil 1 only.
With this technique, the magnetometer signal is zero in the absence of the conductive object and the magnetometer should not be affected by power broadening or non-linearities as long as the field from the eddy currents is smaller than $ \Bsat$
[see Eqs.~(\ref{eq:Jyprime}) and (\ref{eq:Jzprime}) with $\Delta =0$, $B_c=0$ and $B_s = \Bec(\vecro)$].
Furthermore, the signal-to-noise ratio of the measurement will improve by a factor $1/\abs{\alpha}$ if the noise in the magnetometer is proportional to the total signal. This is a dramatic improvement as 
$\abs{\alpha}\approx 10^{-4}$ for our measurements on salt-water.

Figure~\ref{fig:calib}(b) shows three data sets. We see the noisy signal when $\mathbf{B_1}$ is applied. When the opposite magnetic field $\mathbf{B_2}$ is applied, the lock-in outputs change sign. Applying both magnetic fields $10\lr{\mathbf{B_1}+\mathbf{B_2}}\approx 0$ at the same time (and increasing the amplitudes by a factor of 10) gives lock-in outputs close to zero with significantly reduced noise. Coil 1 and 2 are connected to two outputs of the same function generator and the amplitude and phase of the two outputs can be precisely set in order to zero the lock-in outputs.
In Fig.~\ref{fig:calib}(c) we see that the Allan deviation is $\approx 130$ times smaller for integration times $\tau\geq 1$~s when applying both magnetic fields compared to only applying $\mathbf{B_1}$.
Taking the factor of 10 into account, we find an improvement in signal-to-noise ratio of 1300 if detecting a low-conductivity object and a smallest detectable relative signal of 
$\abs{\amin^{\mathrm{diff}}} \approx 1.3\times 10^{-5} $ 
when using the differential technique.

\begin{figure}
\includegraphics[width=0.48\textwidth]{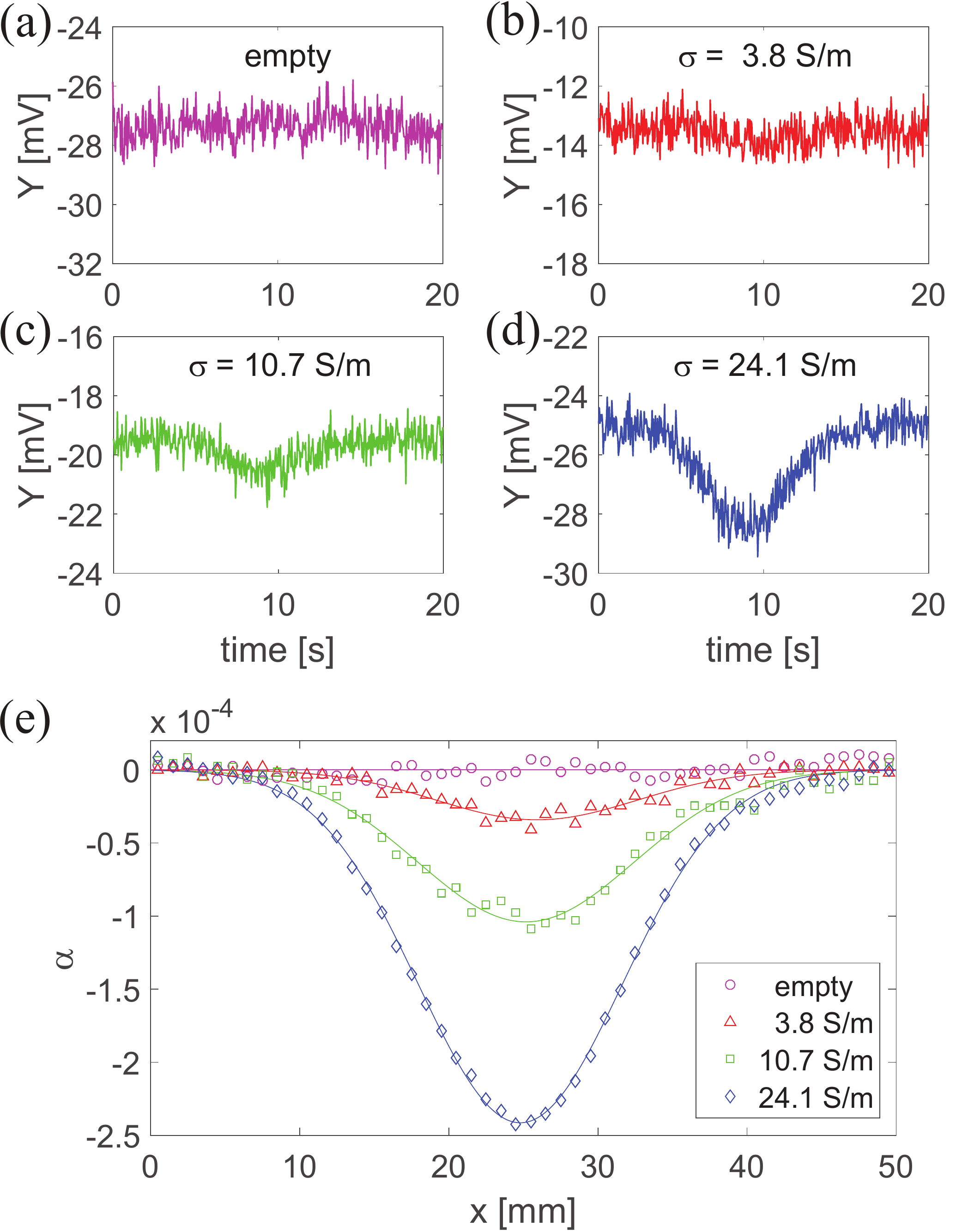}
\caption{(a)--(d): Real-time detection of salt-water. Each data point is integrated for $\tau=40$~ms. 
(e) Relative change in signal using $\approx 20$ averages. Data are binned according to their position with one binned data point per 1~mm.}
\label{fig:saltwater}
\end{figure}

\begin{figure*}[th]
\centering
\includegraphics[width=0.9\textwidth]{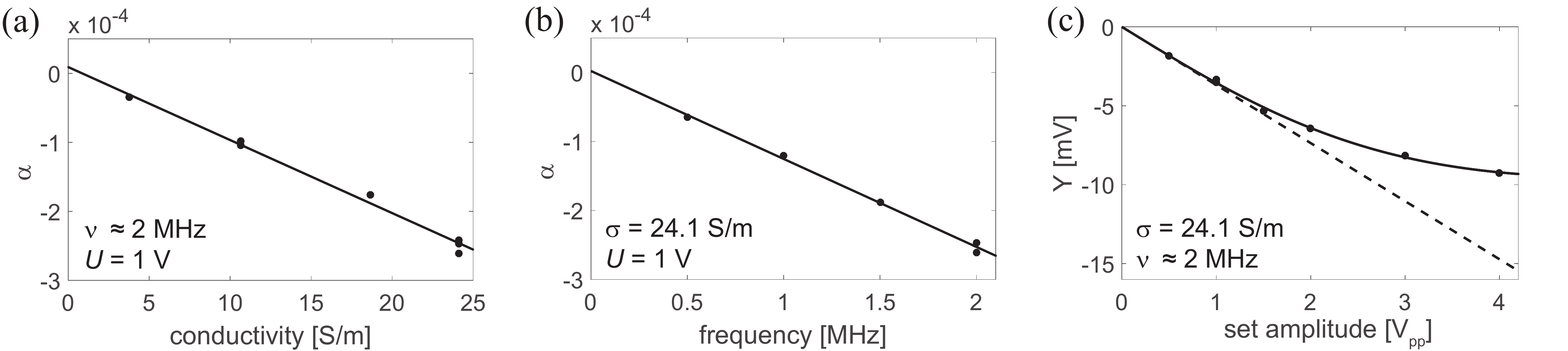}
\caption{(a)--(b) Relative change in signal as a function of conductivity  and applied frequency. Data are shown together with linear fits. (c) Change in signal in mV as a function of the set amplitude $U$ on the function generator connected to the two coils. 
Data are shown together with a fit to the function 
$c \cdot U /\lr{1+\LR{U/\Usat}^2}$ (solid line) and the linear part of the fit function $c \cdot U$ (dashed line).  
A 1~V set amplitude corresponds to the fields $10B_1=-10B_2=45$~nT peak-to-peak amplitude.}
\label{fig:systematicmeas}
\end{figure*}

We now continue with detecting salt-water inside a small container. 
The conductivity of the water can be conveniently varied between 0--24~S/m by changing the concentration of salt. 
Using a motorized translation stage, we scan the container 50~mm in the $x$-direction a few mm above coil 1.
Real-time traces of the $Y$-output when the container is either empty or filled with salt-water with varying conductivity 
are shown in Fig.~\ref{fig:saltwater}(a)--(d). 
With salt-water present, we clearly see a change in the $Y$-output when the container is on top of coil 1 (at the time around 10~s). The $X$-outputs (not shown) are close to zero and do not change during the scan (within the statistical uncertainties).  
In order to reduce noise, the container is scanned $\approx 20$ times over coil 1 and the recorded traces are averaged.
Figure~\ref{fig:saltwater}(e) shows the relative change in signal $\alpha$ for the averaged traces as a function of position. In order to guide the eye and to extract the maximum change in signal, we fit the data with salt-water to a Gaussian function. For the $\sigma = 10.7$~S/m data we have the maximum change 
$\abs{\alpha} = 1.0\cdot 10^{-4}$ which agrees reasonably well with the expected value \cite{Sup}.
The maximum change in signal is plotted in 
Fig.~\ref{fig:systematicmeas}(a) as a function of conductivity and we observe a linear dependence as expected from Eq.~(\ref{eq:BecB1}) 
confirming that the small observed signals are due to the salt-water.
We also vary the applied frequency (while at the same time adjusting the bias field to fulfill the resonance condition $\omega=\gamma B_0$), 
and as shown in Fig.~\ref{fig:systematicmeas}(b) we again observe the expected linear behavior.
Finally we vary the amplitude of the applied field. The signal starts out growing linearly but then some saturation occurs for higher amplitudes
[see Fig.~\ref{fig:systematicmeas}(c)]. 
The data are fitted to the function $c \cdot U /\lr{1+\LR{U/\Usat}^2}$ and we extract the saturation parameter $\Usat = 5.2(2)$~V. This saturation is not expected when using the differential technique. We note that when only $\mathbf{B}_1$ is applied, saturation happens at 10 times lower amplitudes. To avoid issues related to saturation, we used the amplitude $U=1$~V for all other differential measurements (and $U=0.1$~V for all measurements with one coil only).

We emphasize that we detect the salt-water with good signal-to-noise ratio (SNR). We calculate the SNR as the maximum change in signal divided by the standard deviation (found from the data recorded with an empty container).
For the traces in Fig~\ref{fig:saltwater}(b)--(d) we have the SNR of 0.8, 2.5 and 6.1 for the conductivities 3.8, 10.7 and 24.1 S/m. For these traces, the integration time was only 40~ms. 
For the average traces in Fig~\ref{fig:saltwater}(e) we have the SNR of 6.4, 20, and 46. This demonstrates that our setup should be capable of detecting objects with conductivity~$<1$~S/m with a SNR $>1$ and that it should be possible to detect and image biological tissue which has conductivity $\sigma \lesssim 1$~S/m with our optical magnetometer.

In conclusion, we have demonstrated detection of small containers with salt-water with conductivity ranging from 4--24~S/m
using an optical magnetometer and a differential technique which improved the signal-to-noise by more than three orders of magnitude.
Our measurements were performed inside a magnetic shield, however, we expect that the differential technique will yield a larger improvement in unshielded conditions \cite{Deans2018revsciinstr} as there will be more magnetic field noise which can be canceled.
The technique also gives a large improvement when detecting objects with high conductivity (such as metal objects) as long as the detected field from the eddy currents is small compared to the primary field which for instance is the case when the object is far away.
We note that during the preparation of this manuscript similar techniques have been reported and used for imaging of structural defects in metal samples \cite{Bevington2019}.  
By further optimizing the sensitivity and long term stability of our magnetometer we expect that high-resolution imaging of biological tissue will be possible.
This will make optical magnetometers promising candidates for localizing conduction disturbances in the heart allowing for non-invasive diagnostics of heart diseases such as, for example, atrial fibrillation\cite{Marmugi2016scirep}. 

\begin{acknowledgments}
We would like to thank M. A. Skarsfeldt for preparing the salt-water solutions and B. H. Bentzen and S.-P. Olesen for motivating discussions.
This work was supported by 
the Danish Quantum Innovation Center (QUBIZ)/Innovation Fund Denmark, 
the ERC AdG QUANTUM-N
and the EU project MACQSIMAL.
\end{acknowledgments}

%%%%%%%%%%%%%%%%%%%%%%%%%%%%%%%%%%%%%%%%%%%%%%%%

%\bibliography{BIBQuantop5} 

\begin{thebibliography}{10}
\bibitem{NonlinearZeeman}
The non-linear Zeeman effect splits the resonance into $2F=8$ resonances each
  separated by $\delta_{\mathrm{NLZ}}=2\nu_L/\nu_{\mathrm{HFS}}=851$~Hz where
  $\nu_{\mathrm{HFS}}=9.192631770$~GHz is the cesium hyperfine splitting.
\end{thebibliography}
%\bibliographystyle{unsrt} 

%%%%%%%%%%%%%%%%%%%%%%%%%%%%%%%%%%%%%%%%%%%%%%%%

%%%%%%%%%%%%%%%%%%%%%%%%%%%%%%%%%%%%%%%%%%%%%%%%

 \end{document}

\title{Supplemental Material to ``Detection of low-conductivity objects using eddy current measurements with an optical magnetometer"} 

\author{Kasper Jensen}
\thanks{corresponding author\\ 
email: Kasper.Jensen@nottingham.ac.uk}
\affiliation{School of Physics and Astronomy, University of Nottingham, University Park, Nottingham NG7 2RD, England, United Kingdom}
\affiliation{Niels Bohr Institute, University of Copenhagen, Blegdamsvej 17, 2100 Copenhagen, Denmark}

\author{Michael Zugenmaier}
\affiliation{Niels Bohr Institute, University of Copenhagen, Blegdamsvej 17, 2100 Copenhagen, Denmark}

\author{Jens Arnbak}
\affiliation{Niels Bohr Institute, University of Copenhagen, Blegdamsvej 17, 2100 Copenhagen, Denmark}

\author{Hans St{\ae}rkind}
\affiliation{Niels Bohr Institute, University of Copenhagen, Blegdamsvej 17, 2100 Copenhagen, Denmark}

\author{Mikhail~V.~Balabas}
\affiliation{Niels Bohr Institute, University of Copenhagen, Blegdamsvej 17, 2100 Copenhagen, Denmark}
\affiliation{Department of Physics, St Petersburg State University, Universitetskii pr. 28, 198504 Staryi Peterhof, Russia}

\author{Eugene S. Polzik}
\affiliation{Niels Bohr Institute, University of Copenhagen, Blegdamsvej 17, 2100 Copenhagen, Denmark}

\maketitle

\section*{Estimation of the induced magnetic field}
\begin{figure}[bh]
\centering
\includegraphics[width=0.5\textwidth]{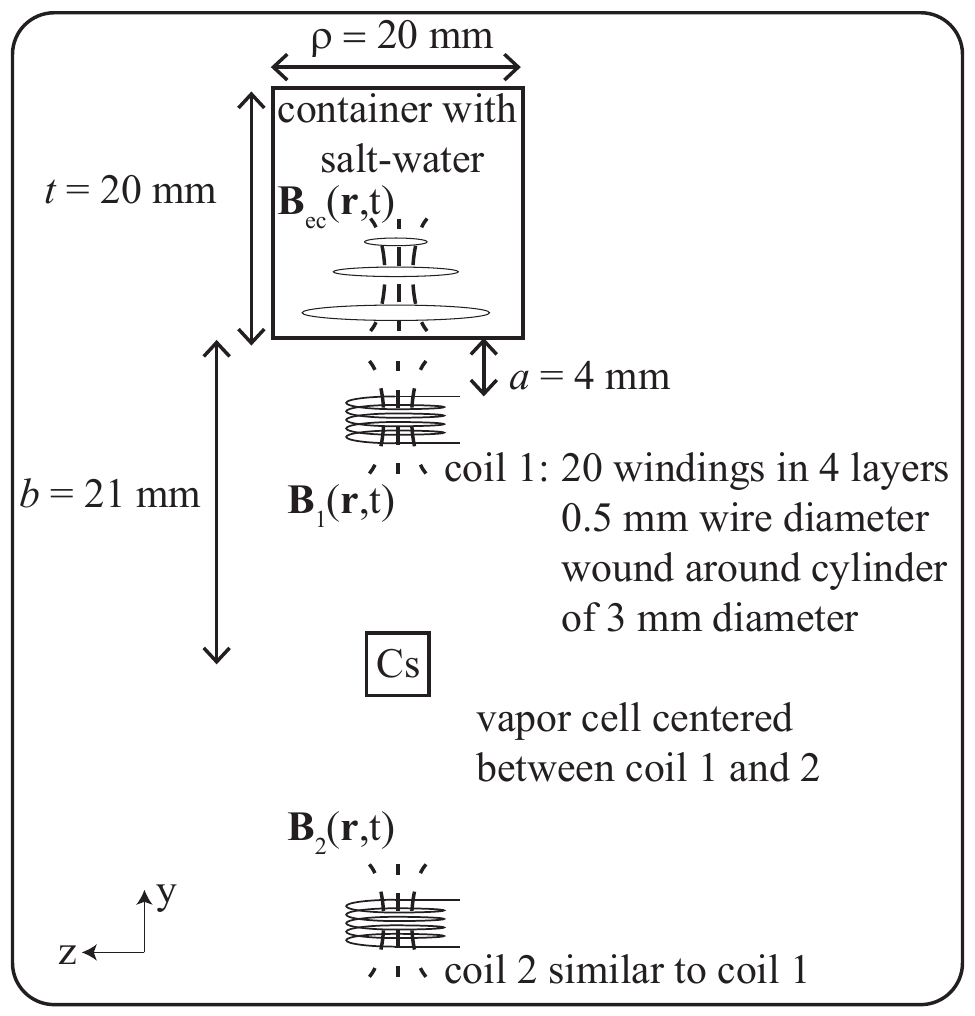}
\caption{Experimental setup with geometrical dimensions.}
\label{fig:setupwithlengths}
\end{figure}

We now estimate the magnitude of the induced magnetic field generated by the eddy currents in the salt-water using a semi-analytical approach based on the calculations in Ref.~\cite{Griffiths1999}.
The real container with salt-water is cubic with $\lr{2~\mathrm{cm}}^3$ inner volume. However, in the calculations below, we will for simplicity assume that the container is a cylinder with radius $\rho$ and height $t$ (also called thickness).
The eddy currents are generated by the primary magnetic field $\mathbf{B_1} $ from coil 1 (see Fig.~\ref{fig:setupwithlengths}). 
The container and the coil are aligned on the same axis in the $y$-direction. 
The coil has multiple windings, but we start by calculating the eddy currents generated by a single winding of radius $r_w$ placed at a distance $a$ from the container.
Assume that an alternating current $I(t)=I_0 \exp (i\omega t)$  with amplitude $I_0$ and frequency $\omega$ is running through the winding.
First, we calculate the induced eddy currents  in a ring of radius $\rho'$ at a distance $a+\tau'$ from the winding, and with radial and axial thickness of $d\rho'$ and $d\tau'$, respectively. 
The induced eddy current is
\begin{equation}
dI_{ec}(\rho') = J d\rho' d\tau',
\end{equation}
where $J$ is the current density. Note that $J=\sigma E$, where  $\sigma$ is the conductivity and $E$ is the magnitude of the electric field. The induced electromotive force $\mathcal{E}$ is firstly given by the integral of the electric field along a closed loop $\mathcal{E} = \oint \mathbf{E} \cdot \mathbf{dl}$. 
Secondly, it is given by
$\mathcal{E}=-d\Phi/dt$, the time derivative of the magnetic flux  $\Phi=\int \mathbf{B_1} \cdot \mathbf{dA}$ of the primary magnetic field $\mathbf{B_1}$
through the enclosed area. 
An analytical expression for the magnetic field $\mathbf{B_1}$ from the winding is given in Ref.~\cite{Simpson2001}.
With the assumption that this magnetic field changes instantaneously with the current in the coil, we find
\begin{align}
& dI_{ec}(\rho') =   \nonumber \\
& \frac{-i\omega\sigma}{\rho'} \exp(i\omega t) \left( \int_0^{\rho'} B_{1}^{(y)}(a+\tau', \rho'')
 \rho''d\rho'' \right)   d\rho' d\tau',
\end{align}
where $B_{1}^{(y)}(a+\tau', \rho'')$ is the amplitude of the 
$y$-component of the magnetic field from the winding at axial and radial distances $a+\tau'$ and $\rho''$, respectively.

To calculate the magnetic field at the center of the vapor cell (distance $b$ from the container) due to the eddy current in a ring of radius $\rho'$ we use the simple expression for the on-axis magnetic field from a current loop 
\begin{equation}
dB^{(y)}_{ec}(b+\tau',0) = \frac{\mu_0 \rho'^2 dI_{ec}(\rho') }{2(\rho'^2 + (b+\tau')^2)^{3/2}}.
\end{equation}
The magnetic field 
at the position $\mathbf{r_0}$ of the center of the vapor cell
from all the eddy currents in the whole salt-water sample is found by integrating the fields from the individual rings
\begin{equation}
B^{(y)}_{ec}(\mathbf{r_0}) = \int_0^t \int_0^{\rho} dB^{(y)}_{ec}(b+\tau',0) d\rho' d\tau'.
\end{equation}
We then add the magnetic fields induced by the individual windings of the coil with their respective distances.

\newpage

The model predicts that the induced field is shifted in phase by $90^{\circ}$ with respect to the primary field, which agrees with our measurements.
We numerically calculate the ratio of the amplitude of the induced magnetic field to  the amplitude of the primary magnetic field taking all the windings of coil~1 into account.
The calculation yields 
\begin{equation}
|B^{(y)}_{ec}(\vecro)|/|B^{(y)}_{1}(\vecro)| = 1.5 \cdot 10^{-4},
\end{equation}
using our experimental parameters 
(see Fig.~\ref{fig:setupwithlengths}), the frequency of 2~MHz, and the conductivity of 10.7~S/m. 
The calculated ratio is $50\%$ higher than the experimentally obtained value.
The main uncertainty on the calculated value stems from the distance between the container and coil 1.
This distance could only be determined with an uncertainty of about 1~mm. 
An increase of the distance between the container and coil~1 by only 1.4~mm would explain the observed deviation of $50\%$ between the calculated and the experimental values.

%\bibliography{BIBQuantop5} 
%\bibliographystyle{unsrt} 